\documentclass[aps,prl,twocolumn,superscriptaddress,groupedaddress]{revtex4}

\usepackage{amssymb}
\usepackage{color,graphicx}
\usepackage{amsmath}
\usepackage{amsbsy}
\usepackage{amsthm}
\usepackage{bbm}
\usepackage{bm}
\usepackage{epsfig}
\usepackage{lscape}
\usepackage{float}
\usepackage{subfigure}

\begin{document}

\title{Narrowing the parameter space of collapse models with ultracold layered force sensors}

\author{A. Vinante}
\email{andrea.vinante@ifn.cnr.it}
\affiliation{Department of Physics and Astronomy, University of Southampton, Southampton SO17 1BJ, United Kingdom}
\affiliation{IFN-CNR and Fondazione Bruno Kessler, I-38123, Trento, Italy}

\author{M.~Carlesso}
\affiliation{Department of Physics, University of Trieste, Strada Costiera 11, 34151 Trieste, Italy}
\affiliation{Istituto Nazionale di Fisica Nucleare, Trieste Section, Via Valerio 2, 34127 Trieste, Italy}

\author{A.~Bassi}
\affiliation{Department of Physics, University of Trieste, Strada Costiera 11, 34151 Trieste, 
Italy}
\affiliation{Istituto Nazionale di Fisica Nucleare, Trieste Section, Via Valerio 2, 34127 Trieste, Italy}

\author{A. Chiasera}
\affiliation{IFN-CNR CSMFO Lab and FBK Photonics Unit, I-38123 Trento, Italy}

\author{S. Varas}
\affiliation{IFN-CNR CSMFO Lab and FBK Photonics Unit, I-38123 Trento, Italy}

\author{P. Falferi}
\affiliation{IFN-CNR and Fondazione Bruno Kessler, I-38123, Trento, Italy}

\author{B. Margesin}
\affiliation{Fondazione Bruno Kessler - CMM, I-38123, Trento, Italy}

\author{R. Mezzena}
\affiliation{Department of Physics, University of Trento, I-38123, Trento, Italy}

\author{H. Ulbricht}
\affiliation{Department of Physics and Astronomy, University of Southampton, Southampton SO17 1BJ, United Kingdom}

\date{\today}

\begin{abstract}
Despite the unquestionable empirical success of quantum theory, witnessed by the recent uprising of quantum technologies, the debate on how to reconcile the theory with the macroscopic classical world is still open. 
Spontaneous collapse models are one of the few testable solutions so far proposed. In particular, the continuous spontaneous localization (CSL) model has become subject of an intense experimental research. Experiments looking for the universal force noise predicted by CSL in ultrasensitive mechanical resonators have recently set the strongest unambiguous bounds on CSL; further improving these experiments by direct reduction of mechanical noise is technically challenging. Here, we implement a recently proposed alternative strategy, that aims at enhancing the CSL noise by exploiting a multilayer test mass attached on a high quality factor microcantilever. The test mass is specifically designed to enhance the effect of CSL noise at the characteristic length $r_\text{\tiny C}=10^{-7}$ m. The measurements are in good agreement with pure thermal motion for temperatures down to $100$ mK. From the absence of excess noise we infer a new bound on the collapse rate at the characteristic length $r_\text{\tiny C}=10^{-7}$ m, which improves over previous mechanical experiments by more than one order of magnitude. Our results are explicitly challenging a well-motivated region of the CSL parameter space proposed by Adler.
\end{abstract}

\maketitle

The question whether the quantum superposition principle remains valid all the way up to the macroscopic domain is still debated. While the widespread belief is that linearity is a fundamental property of nature \cite{zeh,zurek}, over and over this assumption has been questioned\cite{schrodinger,bell,weinberg2,leggett,arndt0}. Spontaneous collapse models \cite{GRW,CSL,adler,DP1,DP2} offer a clear and, under fairly general assumptions \cite{gisin1,gisin2}, unique phenomenology describing the break-down of quantum superpositions when moving towards the macroscopic scale, while preserving the quantum properties of microscopic systems. By construction they are empirically falsifiable \cite{collapse_review2}, and are therefore attracting increasing theoretical and experimental interest \cite{arndt,arndt2,xray,xray1,xray2,adlervinante, bahramiphonons, misra, neutron, tilloy, collett, adler2005, nimmrichter, diosi, vinanteCSL1, vinanteCSL2, lisa, helou, CSLrotational, levZheng, masaki, coldatoms}. 

The general assumption of collapse models is that a universal classical noise drives the state of any material system towards a localized state, even in absence of any measurement process. An inbuilt  amplification  mechanism makes sure that the collapse scales with the size of the system, so that only sufficiently macroscopic objects are effectively localized \cite{collapse_review2}.


In this work, we present a new experimental test of the Continuous Spontaneous Localization (CSL) model \cite{CSL,adler}. In CSL the noise is characterized by two phenomenological parameters: the collapse rate $\lambda$, measuring the strength of the collapse, and a characteristic length $r_\text{\tiny C}$, defining its spatial resolution. The conservative values  $\lambda \simeq  10^{-17}$\,s$^{-1}$ and $r_\text{\tiny C}=10^{-7}$\,m~\cite{GRW,CSL} were initially proposed by Ghirardi {\it et al.} \cite{GRW,CSL} by assuming that the collapse becomes effective at the transition between the mesoscopic and the macroscopic world. A larger value for $\lambda$ has been suggested by Adler \cite{adler}, under the assumption that the collapse is already effective at mesoscopic scale, resulting in $\lambda$ $\sim  10^{9 \pm 2}$ times larger than at $r_\text{\tiny C}=10^{-7}$\,m, and $\sim  10^{11 \pm 2}$ times larger at $r_\text{\tiny C}=10^{-6}$\,m. Moreover, according to Adler, values much larger or smaller of $r_\text{\tiny C}$ are physically less motivated  \cite{adler}. 

The current strongest experimental bounds on the CSL parameters come from noninterferometric tests, which exploit an unavoidable indirect effect of collapse models, namely a tiny violation of the energy conservation \cite{GRW}. 
Relevant examples are spontaneous X-ray emission from Germanium~\cite{xray,xray1,xray2}, spontaneous heating of massive bulk systems \cite{adlervinante,bahramiphonons,misra,neutron,tilloy} or universal force noise on mechanical systems \cite{collett,adler2005, nimmrichter,diosi,vinanteCSL1,vinanteCSL2,lisa, helou,CSLrotational,levZheng, masaki}. Bounds based on the first two effects are already ruling out Adler's parameters, but they can be easily evaded by reasonable assumptions on the spectrum of the CSL noise \cite{cCLSopto,adlervinante}.
Conversely, experiments based on mechanical resonators, with characteristic frequency in the mHz--kHz range, are more robust against changes in the noise properties.

In Ref.~\cite{vinanteCSL2} some of us reported an excess noise in a low temperature cantilever, which could be in principle explained by CSL according to Adler's parameters \cite{adler}. Here, we explicitly test this hypothesis by implementing a novel method to significantly enhance, by almost two orders of magnitude, the CSL noise, thereby circumventing the intrinsic difficulties of a further direct reduction of thermal and background noise in these experiments.
Following Refs.~\cite{nimmrichter, diosi, vinanteCSL1}, the one-sided spectral density of the CSL force noise on the $x$ direction acting on a mass density distribution $\rho(\bm r)$ can be written as:
\begin{equation}\label{SFcsl}
S_{F_\mathrm{CSL}}=\frac{\hbar^2\lambda r_\text{\tiny C}^3}{\pi^{3/2}m_0^2}\int\text{d}\bm q\,q_x^2e^{-q^2 r_\text{\tiny C}^2}|\tilde\rho(\bm q)|^2,
\end{equation}
where $\tilde \rho(\bm q)$ is the Fourier transform of $\rho(\bm r)$ and $m_0$ is the nucleon mass. The effect described by Eq.~\eqref{SFcsl} features a non-trivial dependence on the geometry, and can be enhanced around a given $r_\text{\tiny C}$ by a properly designed multilayered test mass, as discussed in detail in Ref.~\cite{multilayer}. In order to detect the smallest possible CSL effect, one needs to minimize the thermal force noise spectral density $S_{F_\mathrm{th}}=4 k_B T m \omega_0/Q$, which calls for mechanical resonators with low temperature $T$, low frequency $\omega_0$ and high $Q$.


\begin{figure}[!t]
\includegraphics[width=8.6cm]{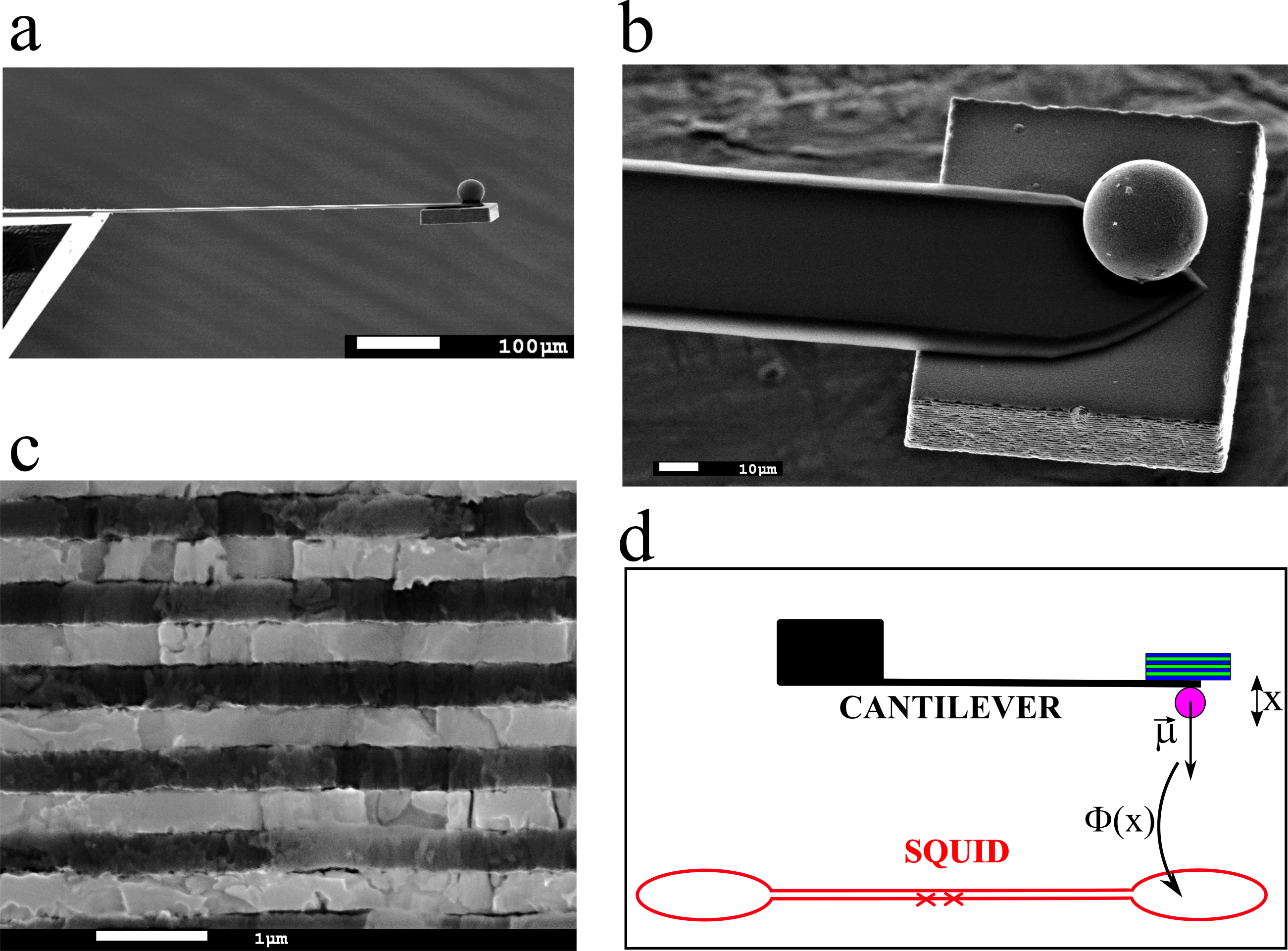} 
\caption{\label{Fig1}Details of the experiment. (a) Low resolution SEM micrograph of the assembled cantilever, with the multilayer test mass and the magnetic microsphere. (b) and (c) SEM micrographs of the multilayer test mass, from top (b) and side (c) view respectively. Here, the alternate layers of WO$_3$ and SiO$_2$ are shown in bright and dark respectively. (d) Simplified scheme of the detection technique, with a gradiometric SQUID magnetometer which detects the variable magnetic field induced by the oscillating ferromagnetic microsphere.}   
\end{figure}

In our experiment, the mechanical sensor is a silicon cantilever (see Fig.~\ref{Fig1}a) of the type developed for atomic force microscopy. The same sensor was used in previous tests of CSL \cite{vinanteCSL2}. A multilayer test mass has been glued on the cantilever end (see Figs.~\ref{Fig1}b,~\ref{Fig1}c). It is a cuboidal structure formed by 47 alternate layers of SiO$_2$ and WO$_3$, fabricated by sputtering. For details on the design and fabrication, see Supplemental Material \cite{suppl}. As described in Ref.~\cite{multilayer}, the multilayer structure enhances the effect of the CSL noise for $r_\text{\tiny C} \lesssim d/3$ where $d$ is the mean layer thickness. The enhancement scales as the density contrast $\Delta \rho=\rho_1-\rho_2$ and the number of layers~\cite{suppl}. In this experiment, we have $\rho_1=7.17 \times 10^3$\,kg/m$^3$ and $\rho_2=2.20 \times 10^3$\,kg/m$^3$, which are respectively the densities of WO$_3$ and SiO$_2$. The mean layer thickness $d=\left( 370 \pm 4 \right)$\,nm, was specifically chosen to maximize the CSL noise enhancement at $r_\text{\tiny C} \approx 10^{-7}$\,m. Based on the measured geometrical parameters, we estimate the value of the multilayer mass $m=\left( 7.1 \pm 0.2 \right) \times 10^{-10}$\,kg.

We attach to the cantilever a second smaller mass, a ferromagnetic microsphere, whose motion is detected by a Superconducting Quantum Interference Device (SQUID) magnetic flux sensor (see Fig.~\ref{Fig1}d) placed at a distance of $\sim 50$\,$\mu$m \cite{vinanteCSL2}. This detection method is very convenient and, owing to the low power dissipated by the SQUID, is compatible with the low temperature regime of the experiment. We notice that the ferromagnetic sphere and the cantilever itself give additional, although smaller, contributions to the CSL force noise, which have been accounted for.

Cantilever and SQUID are enclosed in a mechanically isolated shielded copper box, thermally linked to the mixing chamber plate of a dry dilution refrigerator. The mixing chamber temperature is stabilized by a PID controller. Before performing any measurement we wait for at least two hours to ensure that the temperature is settled, although the thermalization time is expected to be much shorter. We measure resonance frequency and quality factor of the fundamental flexural mode of the cantilever by means of ringdown measurements. The resonance frequency is $f_0=3532.7$\,Hz, while it was measured as $f_0'=8174$\,Hz before attaching the multilayer test mass and with the magnetic sphere already in place \cite{vinanteCSL2}. Accordingly, we use the added mass method to estimate the effective stiffness $k$ of the cantilever with respect to the effective position of the test mass on the cantilever. We obtain the value $k=\left( 0.43 \pm 0.01 \right) $\,N/m. As observed in a previous experiment, the intrinsic quality factor depends slightly on temperature \cite{suppl}, likely due to two-level systems in the silicon cantilever \cite{vinanteCSL2}. The maximum measured quality factor is $Q=\left( 2.83 \pm 0.03 \right) \times 10^6$ at the lowest operation temperature $T=30$\,mK. Remarkably, attaching the large test mass on the cantilever did not spoil the very high $Q$ factor \cite{vinanteCSL2}. This result was crucial to keep a very low thermal noise and was achieved through a careful gluing procedure \cite{suppl}. However, a larger mass implies higher sensitivity to acceleration noise from external vibrations. For this reason we developed a new three-stage mass-spring suspension, improving the isolation at $f=f_0$ by about 40 dB.

\begin{figure}[!t]
\includegraphics[width=8.6cm]{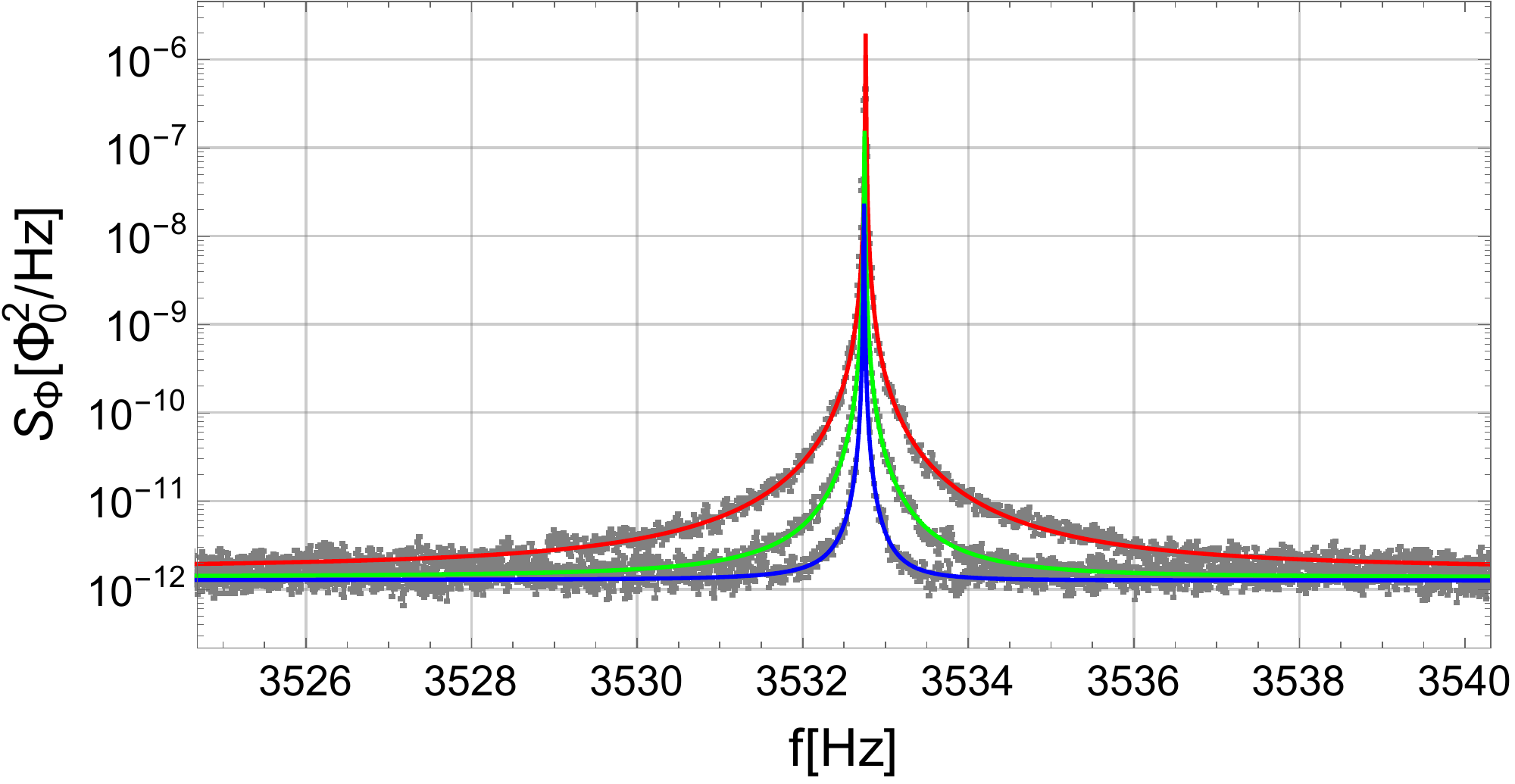}
\caption{\label{fig.2}Representative averaged power spectra of the SQUID flux noise around the cantilever fundamental resonance. The curves refer to the temperatures $T=1000$\,mK, $T=200$\,mK and $T=30$\,mK respectively from the top to the bottom. 
The solid lines are the best fits to the three dataset with Eq.~\eqref{fit}.}  
\end{figure}

At a given temperature $T$, we estimate the power spectral density of the force noise by acquiring and averaging a large number of high resolution periodograms of the SQUID magnetic flux signal. The cantilever motion appears as a resonant peak centered at $f_0$ on top of a white noise floor mainly due to the SQUID imprecision noise. The amplitude of the peak depends on $T$. Some representative averaged spectra are shown in Fig.~\ref{fig.2}. We perform a weighted fit of each spectrum with the theoretical curve \cite{suppl} expressed by:
\begin{equation}
S_\Phi   = A +  \frac{B{f_0 ^4 }+C(f^2-f_1^2)^2}{{\left( {f^2  - f_0 ^2 } \right)^2  + \left( {\frac{{ff_0 }}{{Q' }}} \right)^2 }},
  \label{fit}
\end{equation}
Here, the apparent quality factor $Q'$ takes the place of the intrinsic (or true) $Q$, which is related to the thermal noise. $Q'$ is generally different from $Q$, due to a well understood cold damping effect induced by the SQUID feedback electronics \cite{vinanteCSL2}. The noise parameters $A$ and $C$ and the antiresonance $f_1$ are related with the SQUID noise operated under conventional flux-locked-loop (i.e. negative feedback), and are almost temperature-independent. A full noise model discussing the origin of these terms is discussed in the Supplemental Material \cite{suppl}. The Lorentzian term amplitude $B$ contains the relevant information on the force noise, and can be expressed as:
\begin{equation}
B=\Phi_x^2 \left( \frac{S_{F0}}{k^2} + \frac{{4k_B T}}{{k\omega _0 Q}} \right),
 \label{B}
\end{equation}  
where $\Phi_x=\text{d}\Phi/\text{d} x$ is the magnetomechanical coupling factor which converts a cantilever displacement $x$ into a SQUID magnetic flux $\Phi$, $S_{F0}$ is the spectral density of any nonthermal force noise, and the last term is the thermal noise, which according to the fluctuation-dissipation theorem is proportional to $T/Q$.
The identification of the latter term in Eq.~\eqref{B} allows us to determine $\Phi_x$ and thus to calibrate any non-thermal contribution to $B$.

\begin{figure}[!t]
\includegraphics[width=8.6cm]{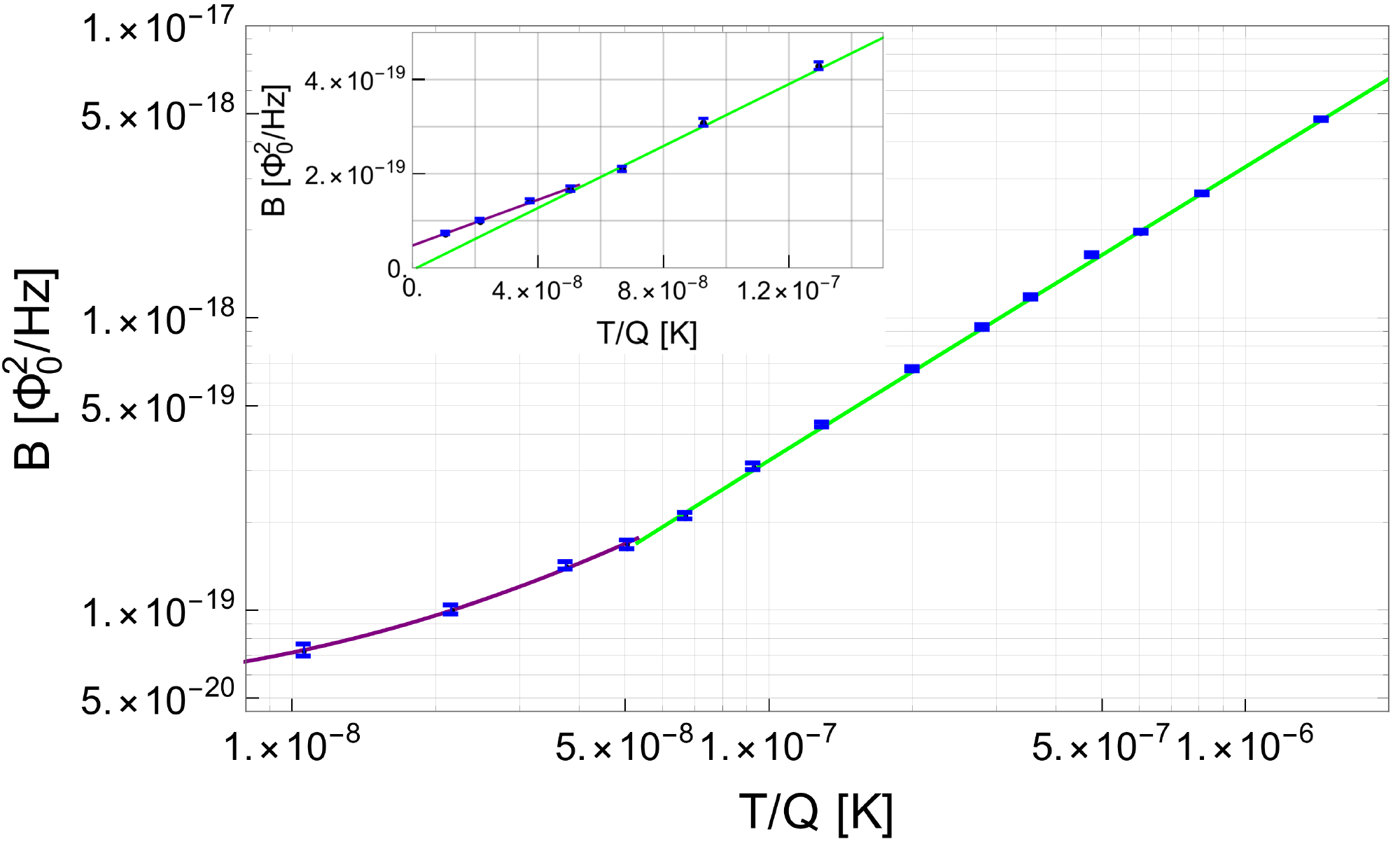}
\caption{\label{data}Measured amplitude of the Lorentzian peak $B$, as function of $T/Q$. The main panel shows all data in Log-Log scale, for better visualization. The inset reports only the low temperature points in linear scale, to underline the crossover between the high temperature and low temperature regimes. The two solid curves represent the linear fits for the points below and above crossover temperature. The linear fit of the data at high temperature is used to bound the CSL noise. }   
\end{figure}

Fig.~\ref{data} shows the measured $B$ as a function of $T/Q$ where $Q=Q(T)$ is measured at the same temperature $T$. The data follow the expected linear behaviour described in Eq.~\eqref{B} down to $T/Q=67$\,nK, which corresponds to $T=100$\,mK. However, the data at lower temperatures (inset of Fig.~\ref{data}) indicate a crossover to a different linear regime, characterized by a lower slope and positive intercept. This behaviour is definitely incompatible with Eq.~\eqref{B} and in particular it cannot be explained by temperature independent noise such as CSL. 

A possible explanation is to assume that at least two dissipation channels are acting on the cantilever motion, one of which is not cooling further below the crossover temperature \cite{vinanteCSL1,usenko}. Formally, we split the dissipation as $1/Q=1/Q_a+1/Q_b$, where $Q_a$ and $Q_b$ are associated to different thermal baths, respectively at the temperatures $T_a$ and $T_b$. In the high temperature limit the system is well thermalized, and $T_a=T_b=T$, where $T$ is the temperature measured by a calibrated thermometer placed on the experimental stage. In the low temperature limit, one of the two temperatures, say $T_a$, saturates to a constant crossover temperature $T_a \approx T_\text{co}$, while the second bath is still well thermalized, $T_b=T$.

Thermal saturations in low temperature systems are typically described by the relation:
\begin{equation}
  T_a=\left( T_\text{co}^n+T^n \right)^{{1}/{n}}.  \label{sat}
\end{equation}
Such a relation is obtained by assuming a steady heat load on the bath at $T_a$ combined with a finite thermal conductance towards the main bath at $T$ varying as $T^{n-1}$ \cite{usenko,pobell}. The most common exponent is $n=4$ and is related to a contact thermal resistance. 
In our case, we suspect that the cantilever motion couples magnetically to dissipating elements located on the SQUID chip, which is expected to saturate in the $50-100$\,mK temperature range. The dissipating elements could be either surface electron spins \cite{naturecomm} or vortices in superconducting films \cite{martinis}. A simple thermal model supporting this possibility is discussed in the Supplemental Material \cite{suppl}. There, we also fit the whole dataset of Fig.~\ref{data} with the combined function $B=B_0+ B_a \left( x^4+x_\text{co}^4 \right)^{{1/4}}+B_b x$, where $x=T/Q$, while $B_0$, $B_a$ and $B_b$ are fitting constants determining respectively the constant contribution and the thermal noise from baths $a$ and $b$. This analysis provides a determination of the crossover at $x_\text{co}=\left( T/Q \right)_\text{co} \approx 53$\,nK which corresponds to $T_\text{co} \approx 85$\,mK. Note that the saturation is effectively very sharp, so that the related excess noise rapidly vanishes for $T \gtrsim T_\text{co}$. However, the data of Fig.~\ref{data} are in principle compatible with other models with larger $n$, and do not allow to make conclusive claims on the actual saturation mechanism. 


In the following, we will make the assumption that the observed crossover is indeed related to a thermal saturation, regardless of its precise physical origin, meaning that for $T<T_{co}$ the data cannot be simply interpreted by Eq.~\eqref{B}. Therefore, to estimate the magnitude of a possible CSL noise effect compatible with the experiment, we restrict our analysis to the range $T \geq 100$\,mK. The restricted dataset follows a linear behaviour remarkably well. A weighted orthogonal linear fit with the function $B_0+B_1 T/Q$, yields the values $B_0=\left(-4.64 \pm 5.31 \right)\times 10^{-21}$\,$\Phi_0^2/$Hz and the slope $B_1= \left( 3.29 \pm 0.03 \right) \times 10^{-12}$\,$\Phi_0^2/ \left( \mathrm{K} \cdot \mathrm{Hz} \right)$. The fact that the intercept is compatible with $0$ is in full agreement with the fluctuation-dissipation theorem, thus indicating that the system is well thermalized in the restricted high temperature range.  According to Eq.~\eqref{B}, the fitting parameters $B_0$ and $B_1$ can be used to estimate the residual non-thermal force noise, which reads:
\begin{equation}
S_{F0}  = \frac{{4k_B k}}{{\omega _0 }}\frac{{B_0 }}{{B_1 }} ,    \label{SF0}
\end{equation}
thus giving $S_{F0} = \left( -1.51 \pm 1.77 \right) \times 10^{-36} $\,N$^2$/Hz.
We use the procedure described in Ref.~\cite{feldman} to determine the upper limit on a strictly positive CSL force noise $S_{F0,\text{\tiny CSL}} \leq 2.07 \times 10^{-36} $\,N$^2$/Hz at the $95 \%$ confidence level. Note that, according to the form of Eq.~\eqref{sat}, any residual effect of saturation in the high temperature data would increase the noise in such a way to increase the value of $B_0$. Therefore, our estimation of CSL noise should be regarded as conservative.
\begin{figure}[!t]
\includegraphics[width=8.6cm]{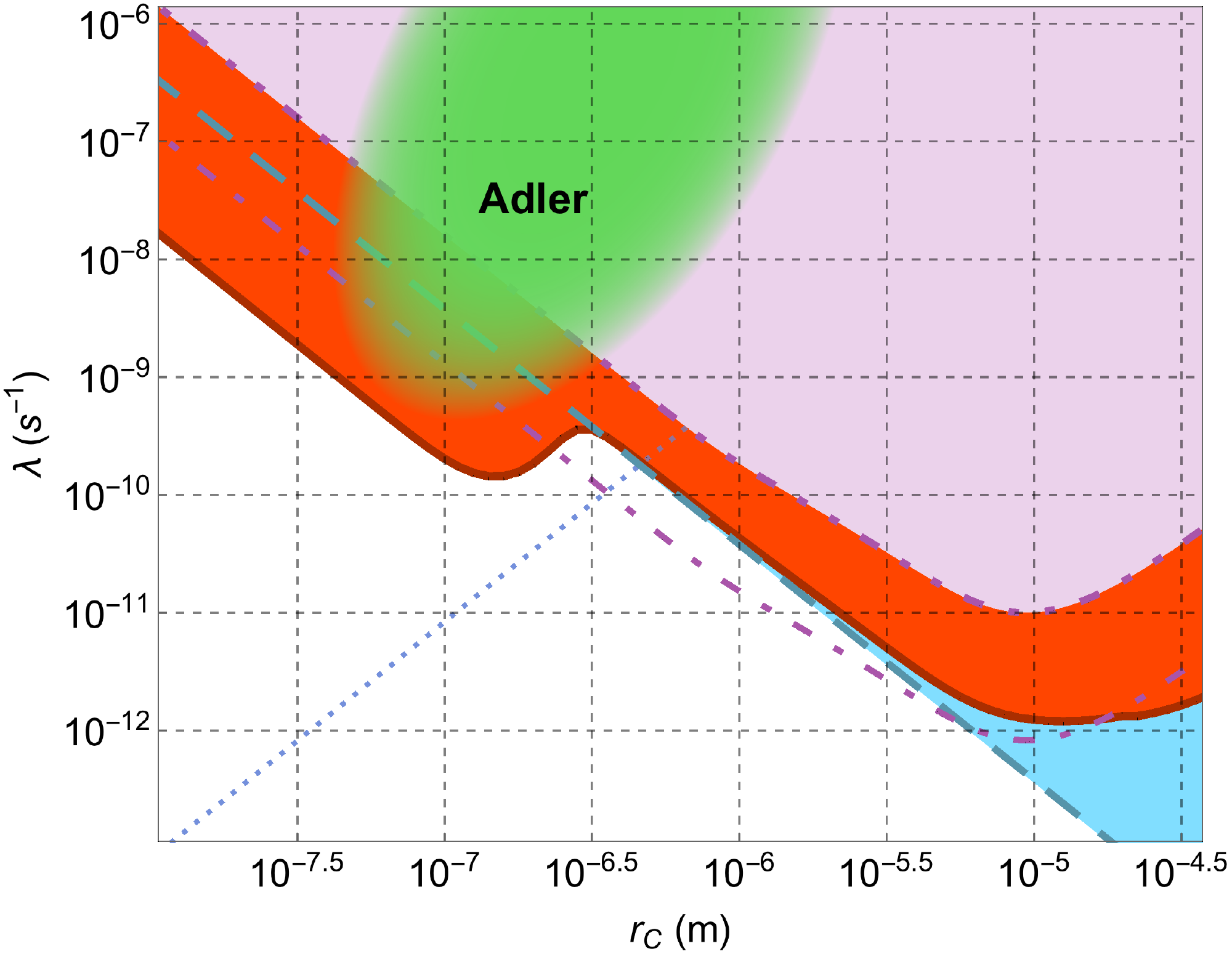}
\caption{\label{fig.4} Exclusion plot for the CSL collapse parameters. Red solid line and shaded area: upper bound and excluded region from the present experiment at the $95\%$ confidence level. Cyan dashed line and shaded area: upper bound and excluded region from LISA Pathfinder \cite{CSLrotational}. Light purple dash-dot-dotted line and shaded area: upper bound and excluded region from a previous cantilever experiment \cite{vinanteCSL2}. Purple dot-dashed line: lower limit of a possible CSL effect from the excess noise observed in the latter experiment \cite{vinanteCSL2}. Blue dotted line: upper bound from X-ray emission from a Germanium sample \cite{xray2}. Since this experiment probes CSL at much higher energies $\sim10^{19}$\,Hz, the upper bound is easily evaded by assuming a spectral cutoff of the CSL noise \cite{cCLSopto}. The green region represents estimations of CSL parameters from Adler, assuming CSL is effective at mesoscopic scale \cite{adler}.}   
\end{figure}

The corresponding upper bound on $\lambda$ is derived taking into account the actual geometry and the materials of the whole mechanical resonator, which is composed by the multilayer mass, the magnetic sphere and the cantilever. The contribution from the multilayer mass is largely dominant at $r_\text{\tiny C}<10^{-7}$\,m, and is responsible for a second minimum of the upper bound at $r_\text{\tiny C} \approx 10^{-7}$\,m \cite{multilayer}. The resulting exclusion plot is shown in Fig.~\ref{fig.4}. The fluctuations in the upper bound due to the uncertainties in the geometry and density of the different subsystems are of the order of the thickness of the curve in Fig.~\ref{fig.4} and cannot be appreciated due to the very compressed logarithmic scale.


The current experiment improves significantly, by almost two orders of magnitude, the previous upper bounds from cantilever experiments at the correlation length $r_\text{\tiny C}=10^{-7}$\,m \cite{vinanteCSL2}, and by more than one order of magnitude the bound from LISA Pathfinder \cite{CSLrotational}. We are thus substantially challenging the parameter region proposed by Adler \cite{adler}. Moreover, the data reveal that the excess noise observed in a previous related experiment \cite{vinanteCSL2} is incompatible with a CSL effect for $r_\text{\tiny C}=10^{-7}$\,m. On the other hand, they do not provide substantial new insight into the origin of that excess noise. Indeed, the absolute value of the excess force noise in the previous experiment, featuring the same cantilever, magnet, and SQUID, is compatible with the error bar of the new experiment. The improved bound on the CSL parameters arises entirely from the largest mass load and the specific multilayer structure. 

We underline that the strong improvement of the bound at $r_\text{\tiny C}=10^{-7}$\,m depends on the peculiar features of the CSL model, which make the force noise sensitive to spatial variations of the test mass internal density \cite{multilayer,diosi2019}. Different localization models may lead to different behaviour. For instance, in the Diosi-Penrose model, the force noise is essentially insensitive to the shape and the spatial distribution of the mass \cite{nimmrichter,diosi}. In principle, specific ad-hoc modifications of the CSL model may lead to a different behaviour as well.

Another point to consider here is that the original estimation of the value of $\lambda$ by Adler was based on very crude assumptions and analysis. Thus, the proposed parameter space represented by the blue region in Fig.~\ref{fig.4} should be taken as indicative \cite{adlerprivate}. In this sense, a further improvement by at least one order of magnitude, possibly with different experimental techniques, may be needed to provide a strong falsification of CSL under Adler's assumptions.

Despite these caveats, our measurements are clearly reducing the probability that CSL effects will be found at $\lambda \gtrsim 10^{-10}$\,Hz. Eventually, one should explore the more conservative framework initially proposed by Ghirardi {\it et al.} \cite{CSL}. In this case, the CSL effects, if existing, could feature a much lower collapse rate $\lambda$. Cantilever experiments may be still improved by 1-2 orders of magnitude, with technological advances in mechanical isolation \cite{tjerk} and a careful characterization of all noise sources. Novel experimental techniques will be needed to fully probe the entire CSL parameter space. Nanomechanical systems at high frequencies \cite{bowen} and levitated microparticles at low frequencies \cite{TEQ,durso,levZheng, maglev} are the most promising routes towards this ambitious goal, together with interferometric techniques on earth \cite{arndt2} and in space \cite{maqro}. 

\begin{acknowledgments}
We gratefully thank S.L. Adler for many stimulating discussions, and N. Bazzanella for technical help. A.B. acknowledges hospitality from the Institute for Advanced Study, Princeton where part of this work was done. We acknowledge financial support from the EU H2020 FET project TEQ (Grant No. 766900), the Leverhulme Trust (RPG-2016-046), the COST Action QTSpace (CA15220), INFN and the Foundational Questions Institute (FQXi). 
The data supporting this study are openly available at https://doi.org/10.5258/SOTON/D1500. 
\end{acknowledgments}

\setcounter{equation}{0}
\setcounter{figure}{0}
\renewcommand{\theequation}{S\arabic{equation}}
\renewcommand{\thefigure}{S\arabic{figure}}

\section{Supplemental Material: Narrowing the parameter space of collapse models with ultracold layered force sensors}

\section{CSL force noise from a multilayer structure}

\begin{figure}[t]
\includegraphics[width=\linewidth]{{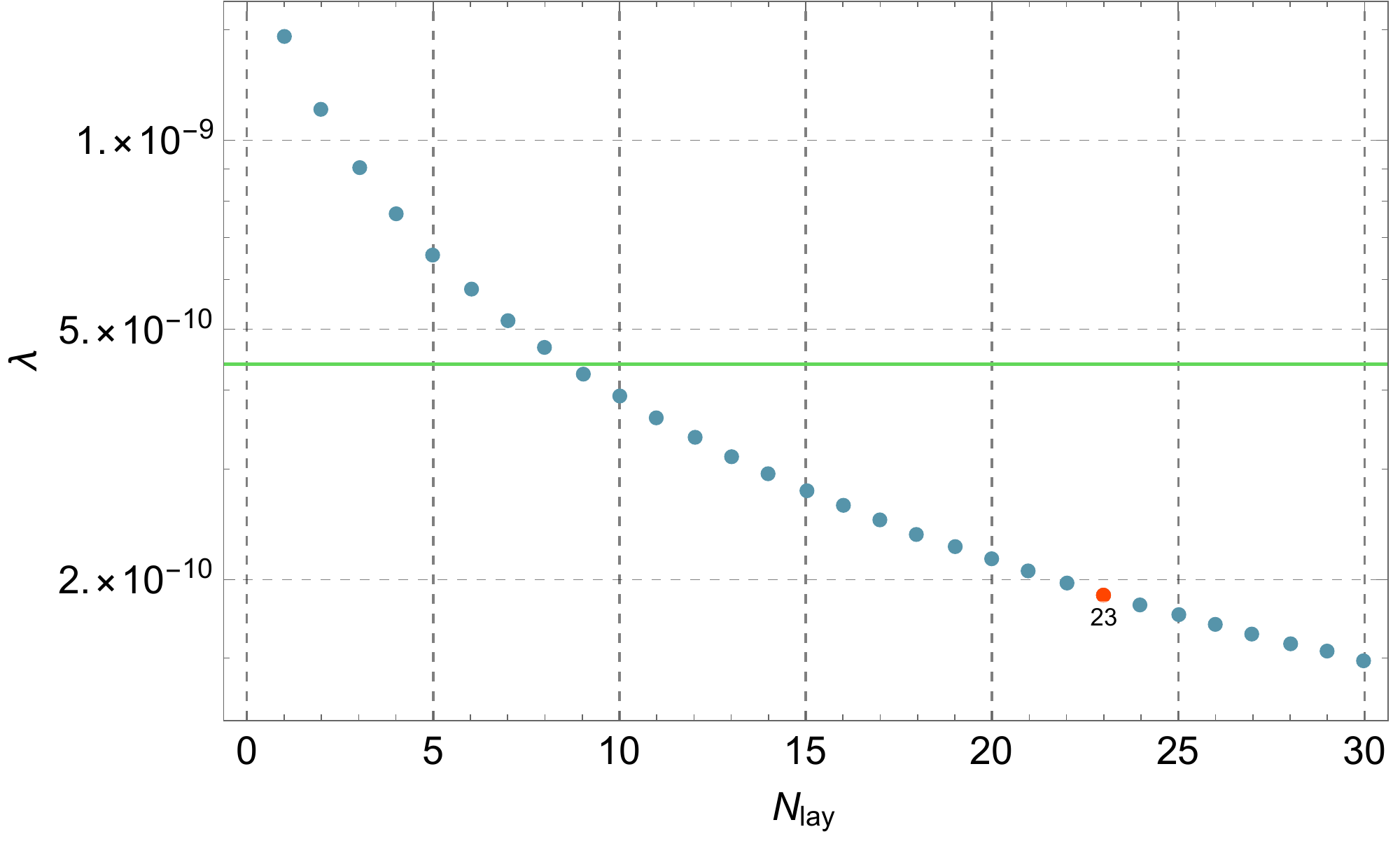}}
\caption{\label{theorymulty} Testable value of $\lambda$ for different values of $N_\text{lay}$ which are shown as blue points. For comparison we report the Adler's lower limit $\lambda=4.4\times 10^{-10}\,$s$^{-1}$ with a green line. The red point corresponds to the value of $N_\text{lay}$ used in our experiment.  }
\end{figure}

The theoretical analysis of the multilayer CSL effect was performed in Ref.~\cite{multilayer} showing that it increases the action of the CSL noise for particular values of $r_\text{\tiny C}$ by suitably choosing the number of layers and their thickness. 
The CSL force noise can be written as
\begin{equation}
S_{F_\mathrm{CSL}}=S^\text{\cite{vinanteCSL2}}_{F_\mathrm{CSL}}+S^\text{multi}_{F_\mathrm{CSL}}+S^\text{interf}_{F_\mathrm{CSL}},
\end{equation}
where $S^\text{\cite{vinanteCSL2}}_{F_\mathrm{CSL}}$ is the CSL force noise due to the cantilever and the ferromagnetic sphere together, whose contributions were studied in \cite{vinanteCSL2}; $S^\text{multi}_{F_\mathrm{CSL}}$ is the contribution due to the multilayer mass and $S^\text{interf}_{F_\mathrm{CSL}}$ is the interference term between the previous two.
Here we focus on $S^\text{multi}_{F_\mathrm{CSL}}$, which reads
\begin{equation}
S^\text{multi}_{F_\mathrm{CSL}}=\frac{\hbar^2\lambda r_\text{\tiny C}^3}{\pi^{3/2}m_0^2}\int\text{d}\bm q\,\sum_{\alpha,\beta}\left(\tilde\rho_\alpha(\bm q)\tilde \rho_\beta^*(\bm q)\right)e^{-\bm q^2r_\text{\tiny C}^2}q_z^2,
\end{equation}
where the sums run over the $N$ layers. As in Ref.~\cite{multilayer}, we assume that there are $N_\text{lay}+1$ layers of density $\rho_1$ and $N_\text{lay}$ layers of density $\rho_2<\rho_1$, where $N=2N_\text{lay}+1$. We assume that all the layers have the same thickness $d$, and that their base is rectangular with length $L_1$ and width $L_2$. Thus, we find that the multilayer contribution to the CSL force noise then is \cite{multilayer}
\begin{equation}
S^\text{multi}_{F_\mathrm{CSL}}=\frac{16 r_\text{\tiny C}^5\lambda}{m_0^2\sqrt{\pi}}J(L_1)J(L_2)I(N_\text{lay},d),
\end{equation}
where we defined
\begin{equation}
J(L_i)=1-e^{-\tfrac{L_i}{4r_\text{\tiny C}}}-\frac{L_i\sqrt{\pi}}{2r_\text{\tiny C}}\text{erf}\left(\frac{L_i}{2r_\text{\tiny C}}\right),
\end{equation}
and 
\begin{equation}
\begin{aligned}
I(N_\text{lay},d)=&\int\text{d}q_z\, e^{-r_\text{\tiny C}^2q_z^2}q_z^2\sec^2(\tfrac{q_z d}{2})\\
&\times\left[\rho_1\sin\left((N_\text{lay}+1)q_z d\right)+\rho_2\sin\left(N_\text{lay}q_z d\right)\right]^2.
\end{aligned}
\end{equation}
In order to design the multilayer structure we numerically investigated the number of layers needed to fully cover the region of parameters proposed by Adler, which goes down to $\lambda=4.4\times10^{-10}\,$s$^{-1}$ at $r_\text{\tiny C}=10^{-7}\,$m. This is done by assuming the following parameters: $L_1=L_2=100\,\mu$m which have been chosen as a factor of $2$ larger than the cantilever width $57\,\mu$m; $\rho_1=7.17\times10^3\,$kg/m$^3$ and $\rho_2=2.2\times10^3\,$kg/m$^3$ which correspond to two materials readily available for the fabrication WO$_3$ and SiO$_2$; and the thickness is chosen equal to $d=320\,$nm which maximizes  the CSL noise precisely at $r_\text{\tiny C}=10^{-7}\,$m \cite{multilayer}. The achievable experimental force noise was set to $S_{F0}=2\times10^{-36}$\,N$^2$/Hz, which is the result from a previous cantilever experiment \cite{vinanteCSL2}.
 The results of the analysis are reported in Fig.~\ref{theorymulty} for different values of $N_\text{lay}$. We find that already for $N_\text{lay}=9$ one can test values of $\lambda$ below $4.4\times10^{-10}\,$s$^{-1}$.  

\section{Sample fabrication}

The WO$_3$/SiO$_2$ multilayer structure was fabricated by RF sputtering technique. The films were deposited on a silicon substrate. Before sputtering, the substrate was spin-coated with a standard $2.1$\,$\mu$m thick positive photoresist, in order to enable releasing of samples at the end of the process. The sputtering deposition was performed by alternating two targets of tungsten oxide (WO$_3$) and silica (SiO$_2$) both with size $15 \times 5$\,cm$^2$. The deposition time necessary to reach the appropriate thickness, are about 2\,h for silica layer and 1\,h and 5\,min for tungsten oxide layers respectively. The residual pressure before the deposition was $2.5 \times 10^{-7}$\,mbar. During the deposition procedure, the substrates were not heated and the temperature of the sample holder during the deposition was $30^{\circ}$\,C. The sputtering occurred with an Ar gas pressure of $5.4 \times 10^{-3}$\,mbar, the applied RF power was $130$\,W and $110$\,W for Silica and Tungsten targets respectively. To monitor the thickness of the layers during the deposition, two quartz microbalances Inficon instruments thickness monitor model SQM-160 faced on the two targets were employed. Thickness monitor was calibrated for the two materials by a long (24 h) deposition process and by directly measuring the thickness of the deposited layer by an m-line apparatus and SEM measurements \cite{chiasera}. The final resolution on the average effective thickness obtained by this quartz microbalance is about 1\,$\AA$. 

After deposition the multilayer is cut into small rectangular chips with approximate dimensions $80 \times 110$\,$\mu$m using a dicing saw, and individual chips are finally released from the substrate using acetone. A selected multilayer chip is manually glued to the commercial cantilever (Nanosensors, type TL-CONT-10) using a small amount ($<1$ picoliter) of Stycast 2850 epoxy. The ferromagnetic microsphere is glued on the other side of the cantilever (see Fig.~1). The microsphere, with radius 15.5\,$\mu$m, is picked from a commercial ferromagnetic powder (Magnequench, type MQP-S-11-9). The dimensions of the multilayer chip, the single layer thickness and the microparticle have been estimated by SEM inspection. The dimensions of cantilever and microsphere have been cross-checked by optical microscopy inspection.

The SQUID is a commercial gradiometric microsusceptometer
composed of two spatially separated Nb loops with radius $10$\,$\mu$m \cite{ketchen}. One loop is used for the cantilever detection, the other is used to apply feedback. The cantilever chip is manually placed above the SQUID (see Fig.~1) with the help of a Macor spacer and firmly held in place by a brass spring. The effective position of the center of the magnetic sphere during the
measurements was about 50\,$\mu$m above the SQUID loop center. The SQUID is read out by a commercial electronics from Magnicon (model XXF-1) normally operated in flux-locked-loop mode, i.e.~with negative feedback.
 
\section{Experimental procedures}

The temperature of the mixing chamber plate is measured by a RuO$_2$ thermometer calibrated against a superconducting reference point device. The temperature was separately cross-checked against a SQUID-based noise thermometer. The overall accuracy of the temperature measurement is estimated as better than $0.5 \%$.

The mechanical quality factor was measured by ringdown measurements. We distinguish the apparent quality factor $Q'$ from the intrinsic quality factor $Q$. During the noise measurements the SQUID is operated in conventional flux-locked-loop mode to provide stable working point and high dynamics. Under these conditions, there is a dynamical change of the quality factor caused by the feedback electronics \cite{vinanteCSL2}. We measure $Q'$ under these conditions. We point out that $Q'$ appears only in the denominator of the resonant terms in Eq.~(2) and due to the fitting procedure its precise value does not affect the estimation of the noise parameter $B$. In contrast, it is very important to determine with good accuracy the intrinsic quality factor $Q$ (which does not include the effects from the feedback electronics) as it determines the strength of thermal noise through Eq.~(3). We measure $Q$ by ringdown measurement by operating the SQUID in open loop, with resolution better than $1\%$. The signal during this measurement is low enough to avoid SQUID nonlinearities. A different procedure involving the measurement of $Q'$ under variable feedback gain and extrapolating to infinite gain \cite{vinanteCSL2} was found to provide results consistent with the first procedure. 

The noise measurements were performed by switching off the pulse tube compressor of the dilution refrigerator, in order to minimize vibrational noise. For each temperature we average a number of periodograms ranging from 40 to 80, corresponding to a total averaging time from $28$ to $56$ minutes.

\section{Noise model}

\begin{figure}[b!]
\includegraphics[width=\linewidth]{{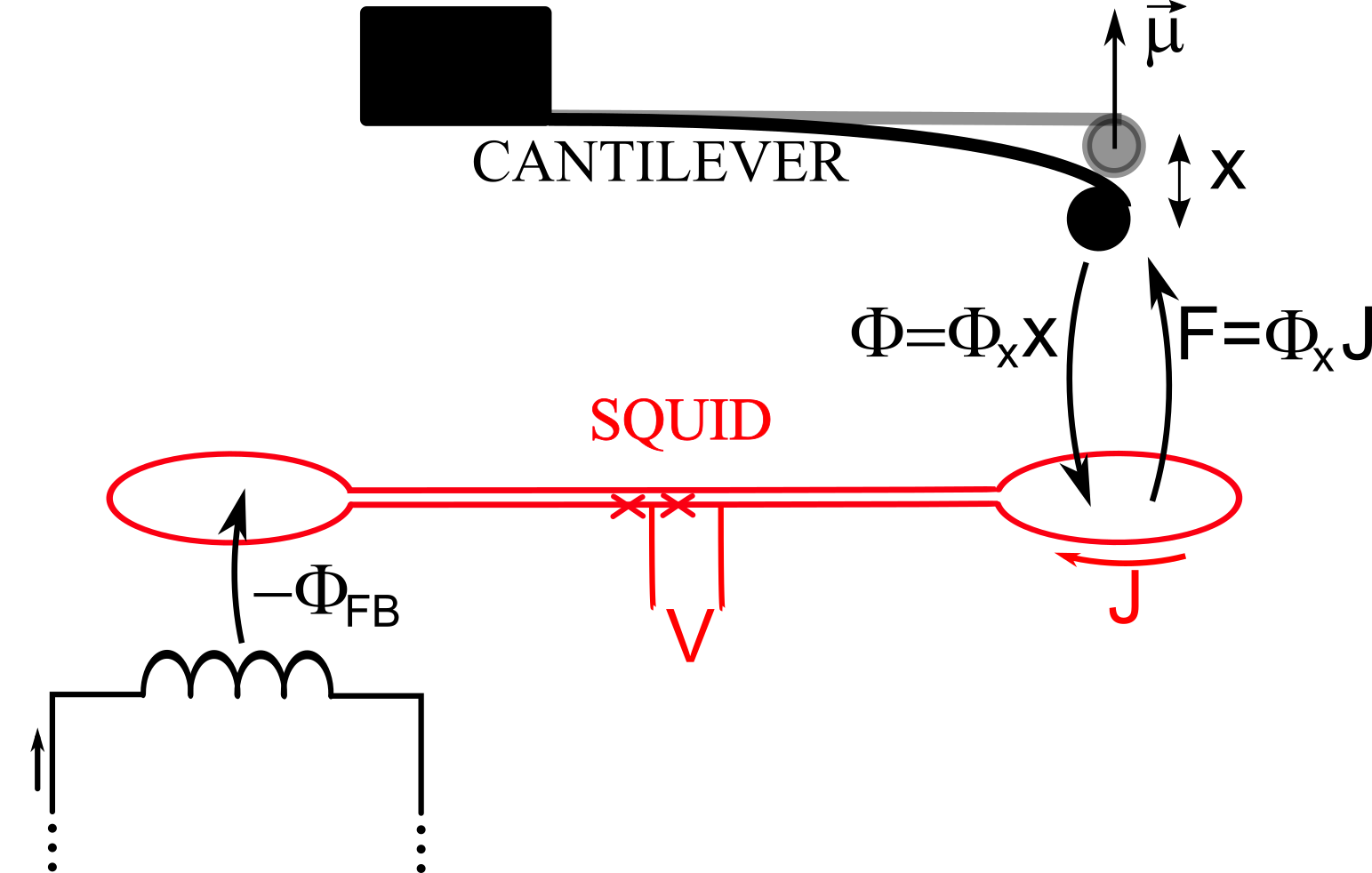}} 
\caption{Simplified scheme of the SQUID-based magnetomechanical detection of the cantilever motion. The magnetomechanical coupling $\Phi_x=d\Phi/dx$ converts a displacement $x$ into a flux $\Phi$ and at the same time it converts a current circulating in the SQUID loop $J$ into a force $F$. The SQUID converts the flux into a voltage $V$ which is read out. The SQUID is operated in flux-locked-loop (i.e. with negative feedback) in order to linearize the SQUID response and stabilize it working point. The feedback flux $-\Phi_{FB}\propto V$ is applied by the electronics through a feedback coil.} \label{figS1}
\end{figure}

The noise model used to fit the spectra is given by Eq.~(2). To understand the origin of the different terms we refer to Fig.~\ref{figS1}.

For our experiment the most relevant term is the one proportional to $B$, which is purely Lorentzian. This is the mechanical noise of the cantilever fundamental mode, driven by thermal force noise and by any additional nonthermal force noise arising for instance by vibrational noise or by CSL. The derivation of the $B$ term in Eq.~(3) follows directly from the displacement spectral density $S_{x}=S_{F} |\chi (\omega)|^2$, where $S_{F}$ is the force noise spectral density and the mechanical susceptibility for a mechanical resonator is:
\begin{equation}
\chi(\omega)=\frac{1}{m \left( -\omega^2+\omega_0^2 +i\frac{\omega \omega_0}{Q'}\right)}.
\end{equation}
Taking into account the magnetomechanical coupling $\Phi_x = d\Phi/dx$, and the cantilever stiffness $k=m\omega_0^2$, one gets immediately the second term in Eq.~(2), with $B$ given by Eq.~(3). The splitting into thermal and nonthermal contribution comes from the expression of thermal force noise $S_{F_\mathrm{th}}=4 k_B T m \omega_0/ Q$.

The origin of the term proportional to $C$ is more subtle. It arises from the fact that the SQUID is operated with negative feedback (see Fig.~\ref{figS1}). This operation mode linearizes the SQUID characteristics and stabilizes the working point. With a negative feedback one effectively measures the flux $\Phi_{FB}$ required to null the total flux in the SQUID. For finite loop gain, it is defined by the condition:
\begin{equation}
\Phi_{FB}= (\Phi-\Phi_{FB}+\Phi_{nA}+\Phi_{nC}) G,  \label{fluxnoise}
\end{equation}
where $G=G(\omega)$ is the open loop gain provided by the feedback electronics, $\Phi$ is the flux applied by the cantilever motion and $\Phi_{nA}$ and $\Phi_{nC}$ are two components of the SQUID flux noise. $\Phi_{nA}$ is the fraction corresponding to a real flux noise, while $\Phi_{nC}$ is the fraction added after the SQUID, referred as a virtual flux noise in the SQUID. This leads to:
\begin{equation}
\Phi_{FB}= (\Phi+\Phi_{nA}+\Phi_{nC}) \frac{G}{1+G}\approx \Phi+\Phi_{nA} + \Phi_{nC}.
\end{equation}
The last approximation is valid for $|G|\gg 1$. For our electronics with bandwidth of 10 MHz we estimate $|G(\omega_0)|\approx 10^3$, so this condition is met.
The total real flux in the SQUID included the feedback (i.e. the error signal), is given by:
\begin{equation}
\Phi_{e}=\Phi+\Phi_{nA} - \Phi_{FB} \approx  \frac{\Phi+\Phi_{nA}}{G}-\Phi_{nC}.
\end{equation}
Note that even in the limit $|G| \to \infty $ the noise term $\Phi_{nC}$ is not nulled by the feedback because it does not correspond to a real flux in the SQUID. Therefore, under high gain and negative feedback a residual real flux noise $-\Phi_{nC}$ is effectively applied. This flux produces a circulating current $J_n=-J_\Phi \Phi_{nC}$ in the SQUID, where $J_\Phi=dJ/d\Phi$ can be either positive or negative, depending on the working point and its order of magnitude is given by $|J_\Phi| \approx 1/L_{SQ}$ \cite{clarke1} where $L_{SQ} \approx 100$ pH is the inductance of our SQUID \cite{ketchen}.
Due to the magnetomechanical coupling, a circulating current $J$ causes a back-action force noise which drives the cantilever $F_n= \Phi_x J_n$, producing an additional noise. It is easy to check that the total effectively measured flux due to $\Phi_{nC}$ will then be:
\begin{equation}
 \Phi'_{nC} = \Phi_{nC} \frac{-\omega^2+\omega_1^2+i \frac{\omega \omega_0}{Q'}}{{-\omega^2+\omega_0^2+i \frac{\omega \omega_0}{Q'}}},  \label{C}
\end{equation}
where:
\begin{equation}
 \omega_1^2=\omega_0^2 \left( 1-\frac{J_\Phi \Phi_x^2}{k} \right)  .  \label{f1}
\end{equation}
The resonance-antiresonance term given by Eq.~\eqref{C} corresponds to the term proportional to $C$ in Eq.~(2) of the paper (where the dissipative term in the antiresonance has been dropped, as it is largely negligible). This term has a very characteristic and very asymmetric shape around resonance, so it can be directly evaluated from a fit of the full PSD. Off resonance, it contributes to a wideband noise.
 
On the other hand, the flux noise component $\Phi_{nA}$ does not produce any back-action, and gives rise to the term proportional to $A$ in Eq.~(2). Note that the SQUID noise terms $A$ and $C$ at kHz frequency are effectively white, and only very weakly dependent on temperature. In fact, the SQUID noise is theoretically limited by Nyquist-Johnson noise in the shunt resistors of the Josephson junctions \cite{clarke2}. However, for this type of SQUID, the temperature of the electrons in shunt resistors is known to saturate at $\approx 0.5$ K due to hot-electron effects \cite{falferiIBM}.

\begin{figure}[b!]
\includegraphics[width=\linewidth]{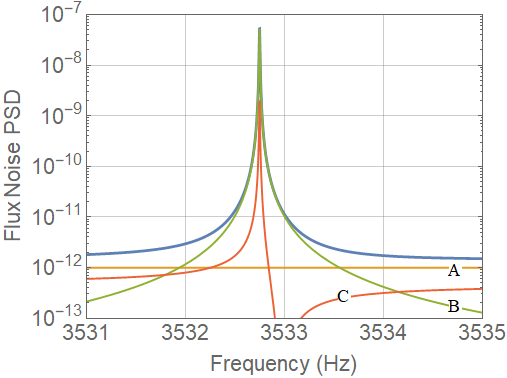}  
\caption{Total flux noise measured by the SQUID, corresponding to the fitting parameters at $T=100$ mK. The three components proportional to $A$, $B$ and $C$ in Eq~(2) are shown. The component relevant to estimate any thermal or external force noise is the pure Lorentzian term proportional to $B$. The term proportional to $C$ is an effective back-action noise caused by the flux-locked-loop operation, and can be efficiently distinguished from the Lorentzian one because of its asymmetry. The term proportional to $A$ is the additive flux noise component.} \label{noiseplot}
\end{figure}

To appreciate the effect of the different noise contributions, we plot in Fig.~\ref{noiseplot} the total noise and the three components for the true fitting parameters in the experiment at $T=100$ mK, which is the lowest temperature considered in the restricted data for the upper limit on CSL. It is apparent that the $C$ term produces a small but significant asymmetry in the total spectrum. This characteristic feature makes the $C$ term well-distinguishable by the fitting procedure.

The data from the fit for the frequency $f_1$ and $f_0$ and the knowledge of the coupling factor $\Phi_x =2.38 \times 10^{7}$\,$\Phi_0/$m, obtained from the thermal noise slope, allow us to estimate the factor $J_\Phi$ in Eq.~\eqref{f1} as $J_\Phi = -2.2 \times 10^{-10}$\,H$^{-1}$, roughly consistent with the expected order of magnitude $|J_\Phi| \approx 1/L_{SQ} \approx 1 \times 10^{-10}$\,H$^{-1}$.

To conclude, we note that SQUID models also predict the existence of an intrinsic noise in the circulating current $J_n$ \cite{clarke2}, which would also contribute to $B$, making our bound on CSL even more conservative. However, this intrinsic back-action noise is estimated to be negligible in our experimental conditions. In fact, it was directly measured in \cite{vinanteCSL2} and found negligible, even if $\Phi_x$ was one order of magnitude larger than in the present experiment.

\section{Data Analysis}
\begin{figure}[b!]
\centering
\includegraphics[width=0.9\linewidth]{{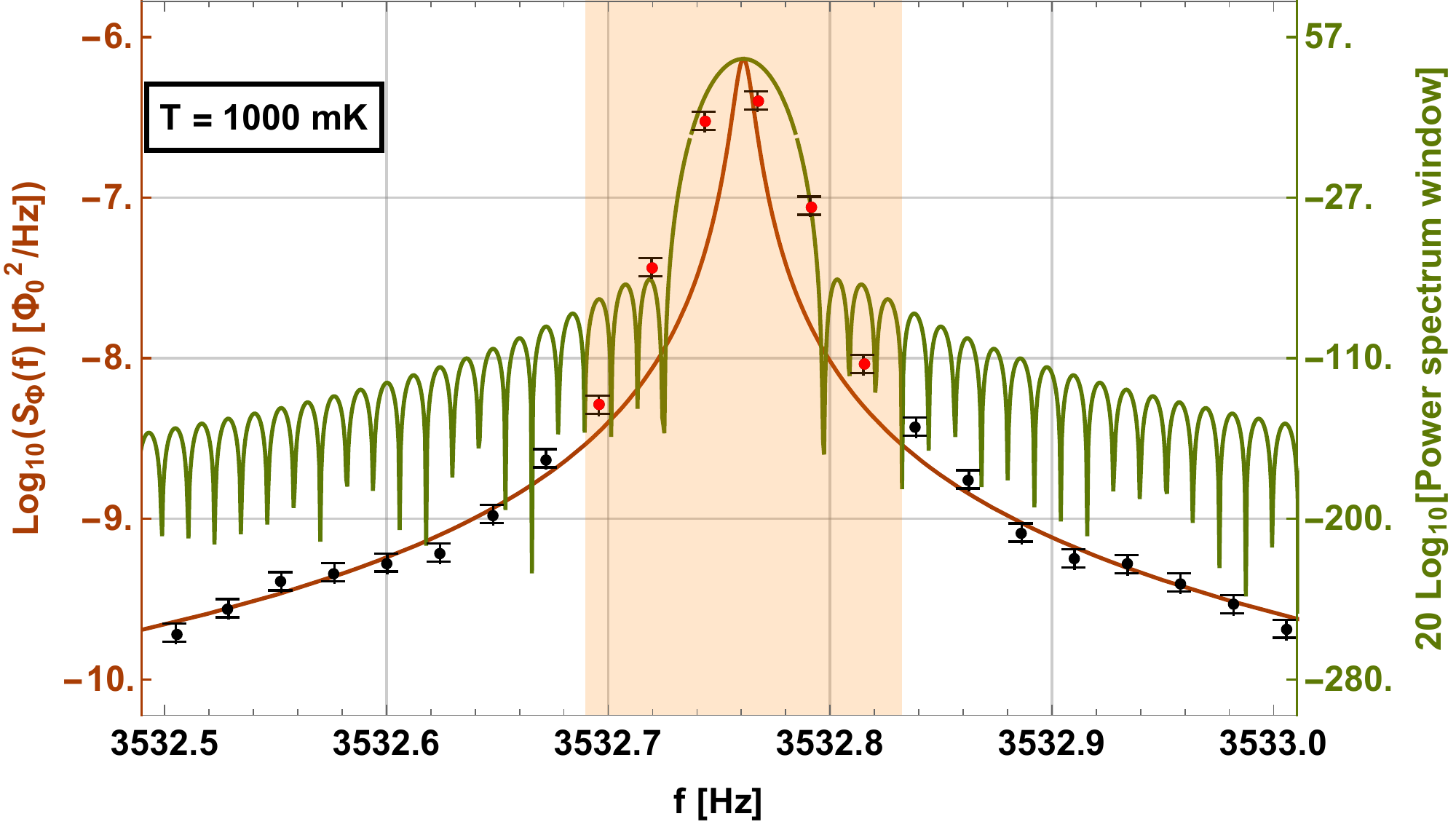}}
\caption{Comparison of the experimental data (red and black dots with corresponding error bars) of the PSD at $T=1000\,$mK with the power spectrum of the Blackman window (green line), which leads to the spectral leakage. The error bars along $f$ are equal to the distance between two subsequent points, here they are not shown to improve the figure readability.
The orange shaded region identifies the frequencies for which the spectrum of the windowing is stronger than -85\,dB. All the points in this region, here highlighted in red, are neglected in the subsequent analysis. Thus, only the black points are used for the fit (brown line).}
\end{figure}
The Power Spectral Density (PSD) $\mathcal S(\omega)$ is experimentally obtained by averaging a number $n_\text{\tiny av}$ of FFT periodograms of the SQUID signal. The sampling frequency is 100\,kHz with a $2^{22}$ samples for each dataset. The latter are weighted with a Blackman window and a FFT is performed. The average of these blocks gives the PSD with a corresponding error which is set to be equal to $\mathcal S(\omega)/\sqrt{n_\text{\tiny av}}$. Finally, the only post-processing of the data is the selection of a window around the peak corresponding to the mechanical motion. Such a window $[3515,3550]\,$Hz is chosen to be the same at all temperatures.

The FFT resolution, despite being quite good, is still larger by a factor $2\sim3$ than the intrinsic width of the peak. This causes a spectral leakage, whose main effect is to distort the PSD around the peak. An effective way to account for this distortion is to remove from the fit all the data points (actually just 6 over more that thousand point used for the fit) that coincide with twice the width of the main lobe of the spectral leakage power. Outside this region, the signal to noise (due to the spectral leakage) ratio improves by -85\,dB. Figure \ref{figS1} shows the comparison between the experimental PSD and the power spectrum of the Blackman window for the dataset at  $T=1000\,$mK.

\begin{figure}[b!]
\includegraphics[width=0.9\linewidth]{{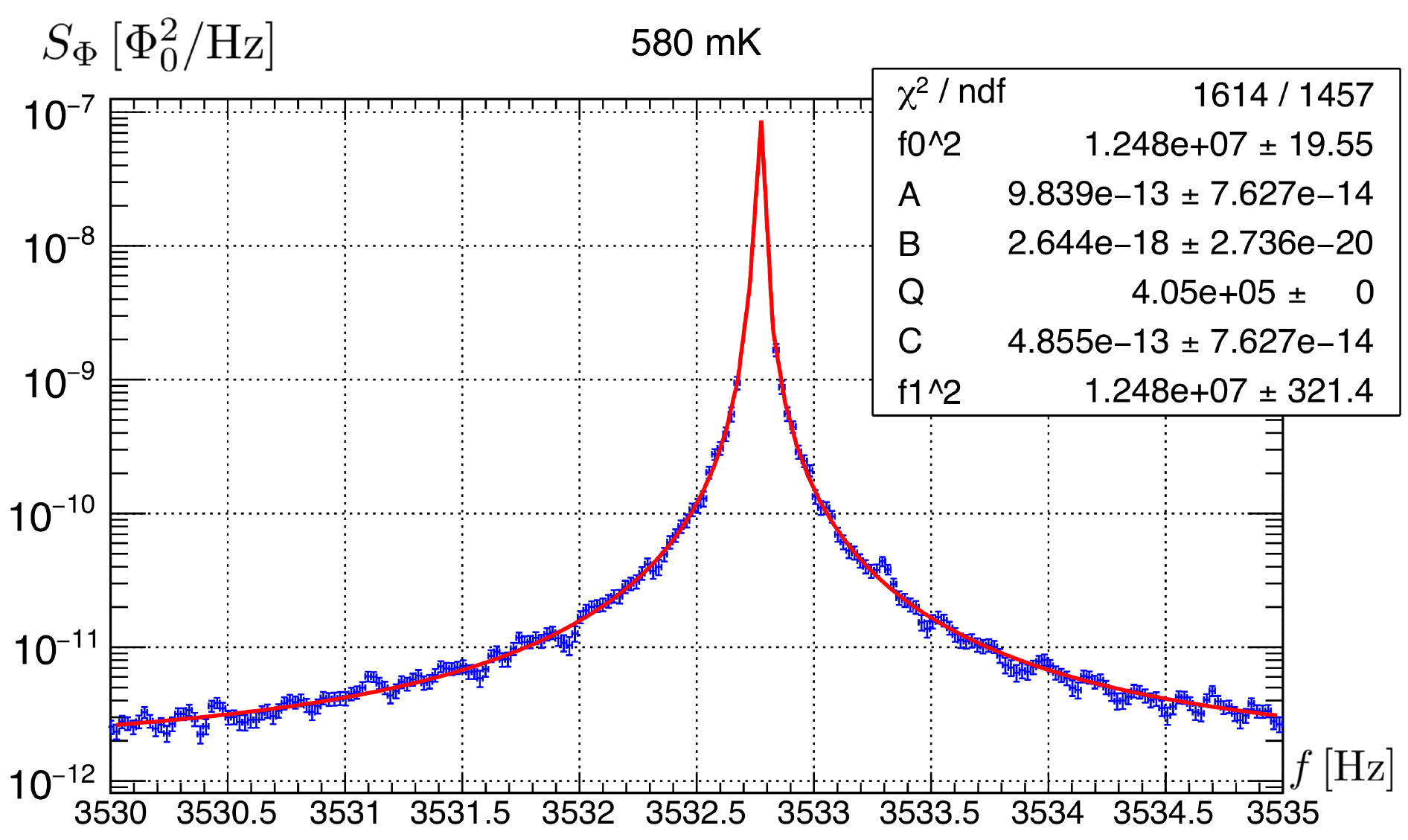}}
\includegraphics[width=0.9\linewidth]{{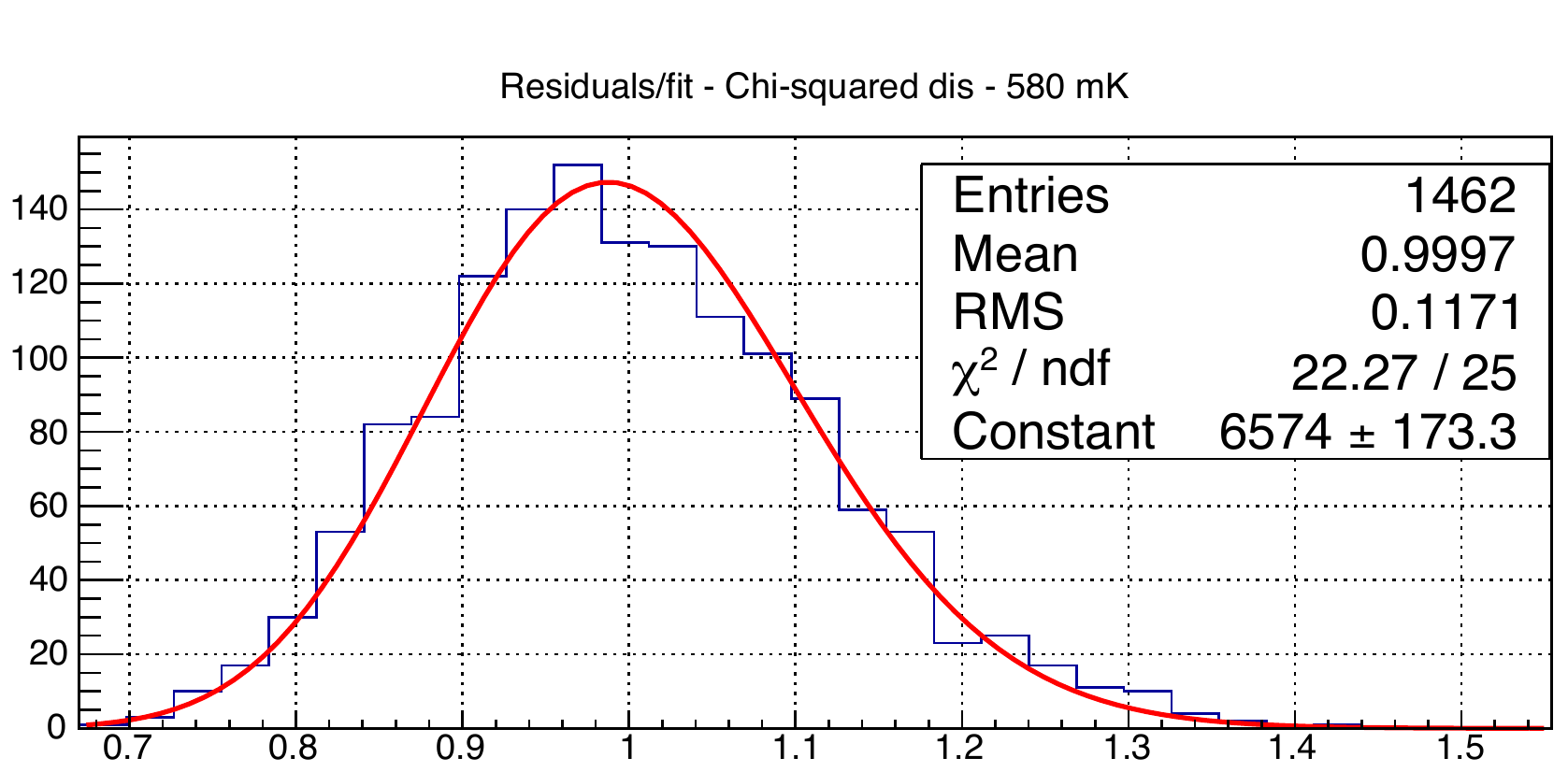}}
\caption{\label{fig580}(\textbf {Top panel}) Example of the experimental data (blue point with corresponding error bars) for $T=580\,$mK, which are fitted with Eq.~(2) (red line). The value of the corresponding $\chi^2$ and of the parameters obtained from the fit are reported in the inset. (\textbf {Bottom panel}) The residual distribution (blue histogram) for $T=580\,$mK, normalized over the fit, is fitted with a $\chi^2$ distribution (red line).  }
\end{figure}
Each averaged PSD at given temperature is then fitted with Eq.~(3) through ROOT. The fit was implemented by applying a recursive fit. In detail, the initial error $\mathcal S(\omega)/\sqrt{n_\text{\tiny av}}$ is substituted at every step with the value of the fitted PSD divided by the square root of $n_\text{\tiny av}$: $\mathcal S(\omega)/\sqrt{n_\text{\tiny av}}\to\mathcal S_\text{\tiny fit}(\omega)/\sqrt{n_\text{\tiny av}}$. Then the recursive fit is performed until the value of $\chi^2/\text{d.o.f.}$ approaches an asymptotic value within $0.01\%$. For every fit, the fixed parameters are $n_\text{\tiny av}$ and  $Q'$, while $A$, $B$, $C$, $f_0$ and $f_1$ are the free parameters of the fit. An example of the fit is shown in the top panel of Fig.~\ref{fig580}, where the data for $T=580\,$mK are analyzed.

Once the recursive fit at the asymptotic value of $\chi^2/\text{d.o.f.}$ near to 1 is found, we implement an additional quality check. This is the fit of the normalized residuals distribution with that of a $\chi^2$-distribution with $2n_\text{\tiny av}$ degrees of freedom. This is what one expects from the error distribution of a power spectrum. Here, we applied the Freedman-Diaconis rule \cite{freedman} to compute the width of the histogram bins.

\begin{figure}[t!]
\includegraphics[width=0.9\linewidth]{{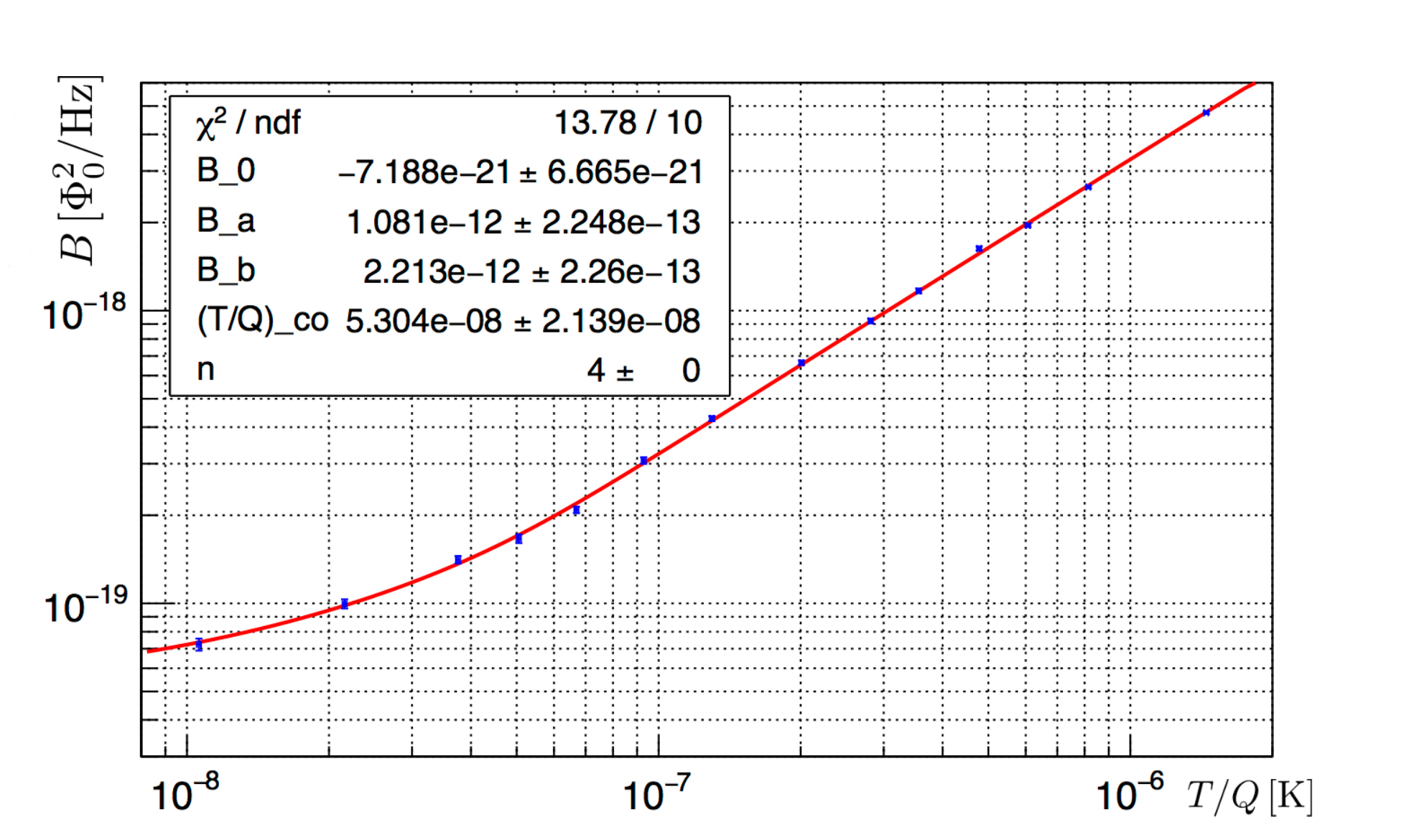}}
\includegraphics[width=0.9\linewidth]{{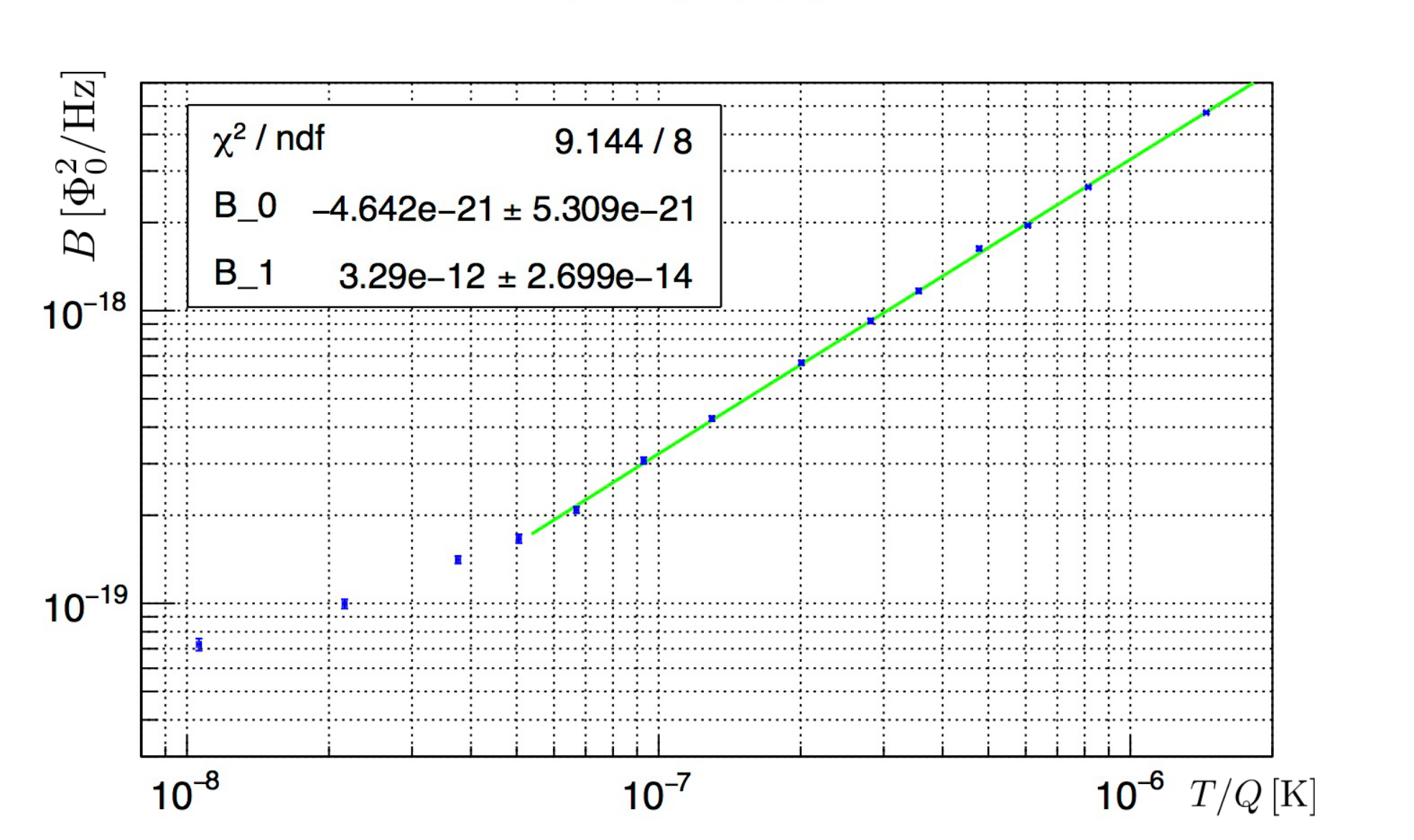}}
\caption{\label{fig.fits} Experimental data (blue point with corresponding error bars) for the values of $B$ at different values of $T/Q$ compared to two different fits. (\textbf {Top panel}) Fit with Eq.~\eqref{eqBlinsat} (red line), which corresponds to a thermal saturation at low temperatures. The value of the corresponding $\chi^2$ and of the parameters obtained from the fit are reported in the inset. (\textbf {Bottom panel}) High temperature fit with Eq.~\eqref{fitlinear} (green line). The value of the corresponding $\chi^2$ and of the parameters obtained from the fit are reported in the inset. }
\end{figure}

The resulting values of $B$ are fitted against $T/Q$, which is the temperature dependent quantity appearing in a thermal force noise $S_{F_\mathrm{th}}=4k_B T m \omega_0 /Q$. Fig.~\ref{fig.fits} reports the corresponding experimental data and the functions used for their fit. Although the overall trend of $B$ is proportional to $T/Q$, the data suffer of a saturation behaviour at low temperatures. This can be fitted with the following function
\begin{equation}\label{eqBlinsat}
B=B_0+B_a \left[\left(\tfrac{T}{Q}\right)^n+\left(\tfrac{T_{co}}{Q}\right)^n \right]^{1/n} + B_b \tfrac{T}{Q} ,
\end{equation}
where the term proportional to $B_a$ describes the deviation from the linear behaviour through the saturation at temperatures below $T_{co}$. In particular, the choice of $n=4$ corresponds to a Kapitza contact thermal resistance as dominating thermalization path, which is the saturation process we expect. In the following we will consider  $n=4$. In the high temperature limit, Eq.~\eqref{eqBlinsat} reduces to
\begin{equation}\label{eqBlinsathigh}
B=B_0+(B_a+B_b) \tfrac{T}{Q}+\mathcal O(\tfrac{1}{T^3}),
\end{equation}
which corresponds to the linear high-temperature behaviour described by the classical fluctuation dissipation theorem. Here, $B_0$ will be the non-thermal contribution, that could be possibly produced by the CSL mechanism.

Since the precise thermal saturation mechanism at work is not fully understood, one is forced to neglect the low temperature data and restrict the analysis to high temperatures. The distinction between low and high temperatures is made by considering the fit of the experimental data with Eq.~\eqref{eqBlinsat}, which gives a value for $T_{co}/Q=(5.30\pm2.14)\times 10^{-8}\,$K corresponding to a temperature just above $T=85$\,mK. Thus, all the data corresponding to $T=100$\,mK and above are considered as high-temperature data, which will be used for a linear fit using
\begin{equation}\label{fitlinear}
B=B_0+B_1{T/Q}.
\end{equation}
This results to $B_0=(-4.64\pm5.31)\times10^{-21}\,\Phi_0^2/$Hz and $B_1=(3.29\pm0.03)\times10^{-12}\,\Phi_0^2/$(Hz\,K) with a $\chi^2/\text{ndf}=9.144/8$. The high-temperature linear fit and the one with the saturation at low temperatures are reported in Fig.~\ref{fig.fits}. For the high-temperature data we also considered a quadratic fit with the function $B=B_0+B_1{T/Q}+B_2({T/Q})^2$, which gives a slightly higher value of $\chi^2/\text{ndf}=8.196/7$. This confirms that nonlinear terms are not statistically significant and that the data follow a pure linear behaviour in full agreement with Eq.~(3).

The residual non-thermal force noise can be computed through Eq.~(5), which gives a value $S_{F0}=(-1.51\pm1.44)\times10^{-36}$\,N$^2$/Hz that is compatible with zero.
The corresponding positive upper bound can be obtained following the methods in \cite{feldman}, which is the technique used in high energy physics to determine upper bounds on the positive rate of rare events. A typical example where such technique needs to be applied is the following. Consider a Gaussian distribution, whose mean $\bar z_0$ is constrained to be non-negative while its measured mean $z_0$ is negative but compatible with zero in few standard deviations. Clearly, the whole confidence interval on $\bar z_0$ is also constrained to be non-negative: its lower bound will be zero while its upper bound will depend on the precise value of the measured mean $z_0$ and the desired confidence level. Ref.~\cite{feldman} prescribes how to determine such a confidence interval according to classical statistics.
Applying this prescription to $S_{F0}$, which is expected to be non-negative since it is a power spectrum, 
we find that its upper bound at the 95\% of confidence level $S_{F0}^\text{\tiny upper}=2.07 \times10^{-36}$\,N$^2$/Hz.

The theoretical CSL-induced PSD, expressed by Eq.~(1), can be computed numerically as in \cite{multilayer}. The mass distribution that has been taken into account includes the cantilever, the multilayer mass and the sphere. The cantilever has density $\rho_C=2.33\times 10^3$\,kg/m$^3$ and dimensions $450\times57\times2.5$\,$\mu$m$^3$.  The ferromagnetic sphere has density $\rho_S=7.43\times 10^3$\,kg/m$^3$ and radius $R=15.5\,\mu$m. Finally, the multilayer mass has a basis of dimensions $113\times82\,\mu$m$^2$ and it is made of 47 alternate layers of thickness $370 \pm 4$\,nm. 24 of them have density $\rho_A=7.17\times10^3$\,kg/m$^3$, while the remaining 23 have $\rho_B=2.2\times 10^3$\,kg/m$^3$. By comparing the numerically computed CSL force noise $S_{F_{CSL}}$ with the experimental upper value of the residual force noise $S_{F0}^\text{\tiny upper}$, one obtains the experimental upper bound on the CSL parameters, which is shown in Fig.~4.

\section{Two-bath model}

\begin{figure}[b!]
\includegraphics[width=0.9\linewidth]{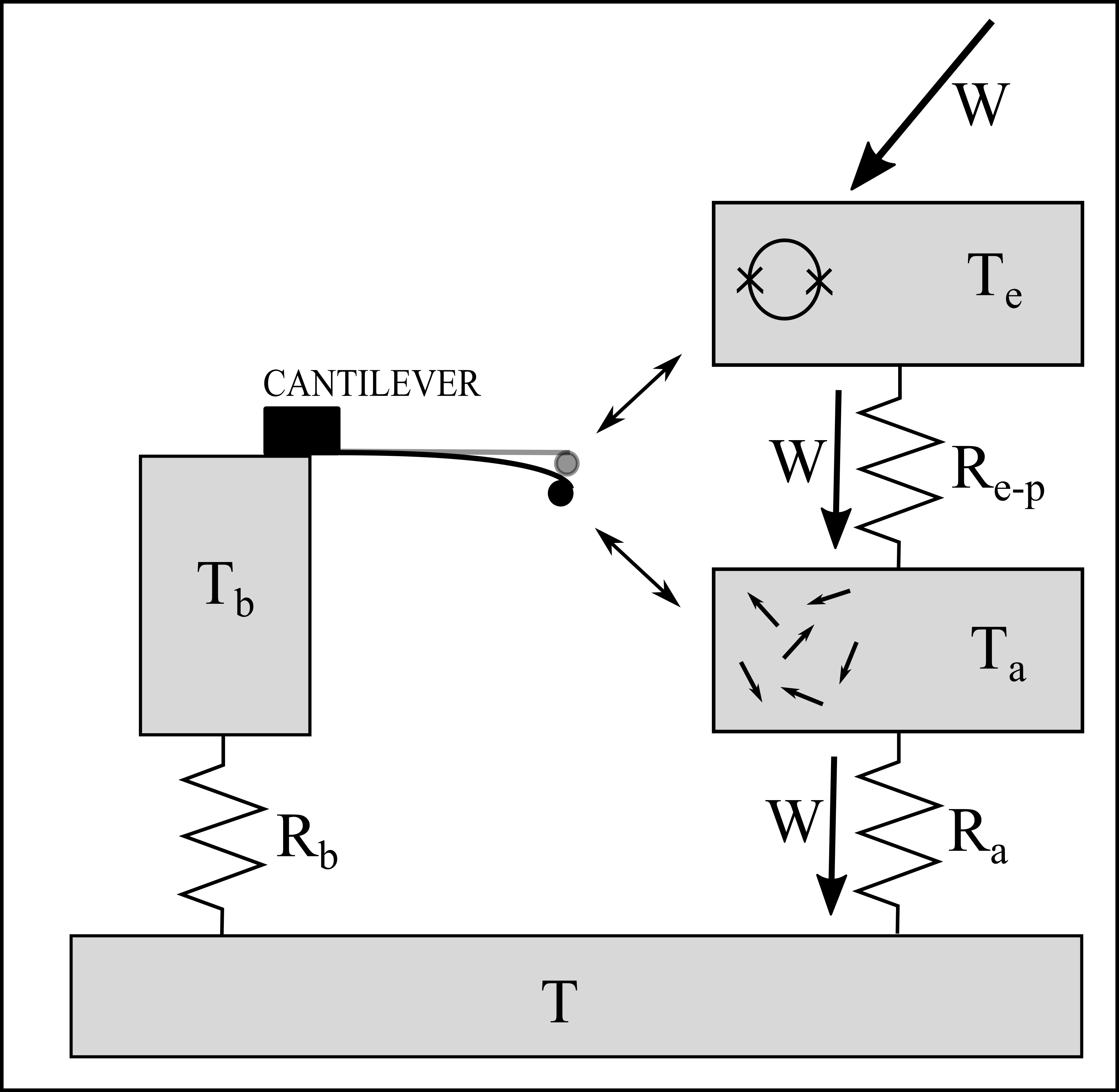}  
\caption{Two-bath thermal model. The cantilever motion features at least two dissipation channels: intrinsic mechanical dissipation in the cantilever (at temperature $T_b$) and noncontact magnetic losses in the SQUID chip which is positioned $50$ $\mu$m underneath the cantilever (at temperature $T_a$). The latter could be due for instance to a bath of paramagnetic spins. The two baths are coupled to the experimental stage on the dilution refrigerator at temperature $T$ (where a thermometer is placed) through different thermal resistances $R_a$ and $R_b$. As there is no external power nominally dissipated in the cantilever, we can reasonably expect that $T_b\approx T$. In contrast, the temperature of the SQUID chip $T_a$ is certainly higher, due to a steady power $W\approx 0.5$ nW dissipated in the SQUID shunt resistors. The power $W$ maintains the electrons in the resistors at a temperature $T_e \approx 500$ mK, due to weak electron-phonon coupling $R_{e-p}$, and the SQUID chip itself at a temperature $T_a>T$, due to the contact thermal resistance $R_a$.} \label{thermalscheme}
\end{figure}

As discussed in the paper, it is possible to interpret the crossover in the data of Fig.~3 in terms of a two bath model. We assume that there are at least two dissipation channels to two different thermal baths, and that one of them decouples from the main bath at temperature $T$ below some crossover temperature $T_{co}$. 

\begin{figure}[b!]
\includegraphics[width=\linewidth]{{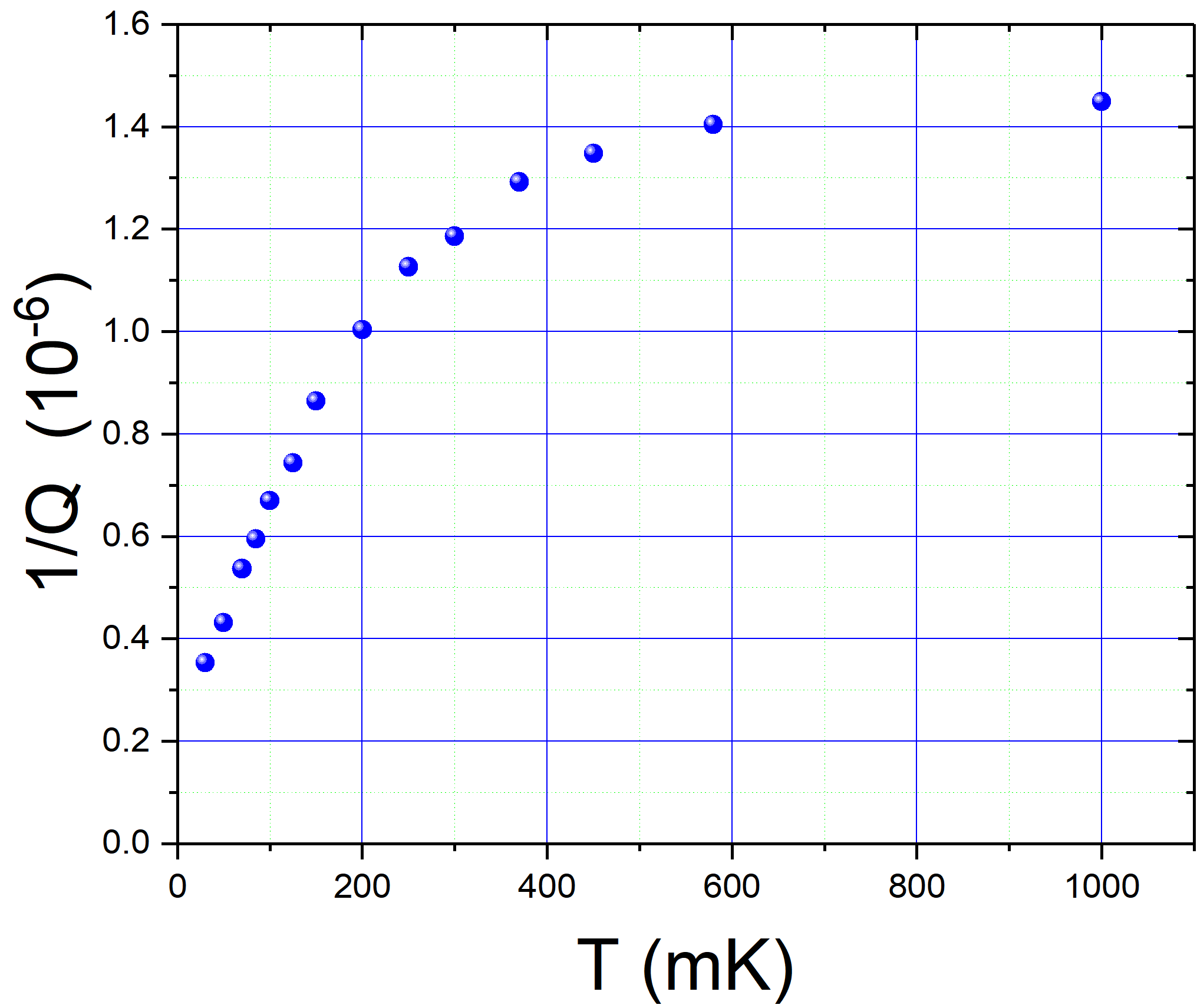}} 
\caption{Dissipation $1/Q$ as a function of temperature $T$. } \label{QvsT}
\end{figure}

For our setup we have actually evidence of at least two dissipation channels, as depicted in the scheme in Fig.~\ref{thermalscheme}. To discuss this point we refer to Fig.~\ref{QvsT}, which shows the dissipation $1/Q$ as function of $T$.
Intrinsic losses due to two-level systems on the cantilever surface are likely explaining the linear dependence of $Q$ at low temperature, with a plateau at $T>500$ mK. A similar behaviour has been observed in past experiments \cite{vinanteCSL2}. This effect is observed down to the lowest temperatures of $30$ mK, indicating that the cantilever structure is always well thermalized, i.e. $T_b=T$. 
A second dissipation channel could arise from non-contact losses in the SQUID chip, due to the magnetic field produced by the oscillating magnetic particle. Possible dissipation mechanisms are the dragging of fluxons on the superconducting niobium film which is patterned to realize the SQUID loops \cite{martinis}, or a layer of paramagnetic spins on the silicon chip surface \cite{naturecomm}. This second dissipation channel could explain the residual dissipation at zero temperature visible in Fig.~\ref{QvsT}. We have found variations of this residual $1/Q$ with magnitude of order $10^{-7}$ by varying the precise position of the cantilever above the magnet in separate runs. This strongly suggests that a noncontact dissipation does indeed exist. The relevant temperature for noncontact losses is that of the SQUID silicon chip, which we define as $T_a$, as shown in  Fig.~\ref{thermalscheme}.

When the SQUID is working, a steady power $W$ is dissipated in its shunt resistors, with $W\approx R I^2 \approx 0.5$\,nW, where $R \approx 5$\,$\Omega$ is the SQUID parallel shunt resistance, and $I\approx 10$\,$\mu$A is the dc bias current used in this experiment. As a first consequence, the temperature of the electrons in the shunt resistors saturates at about $T_e\approx 500$\,mK due to the so-called hot electron effect \cite{wellstood}. Namely, the very weak electron-phonon interaction determines an effective thermal resistance, and the electron temperature follows Eq.~(4) with $n=5$ \cite{wellstood, hotelectron}. This mechanism is well-understood and cannot explain the observed saturation at below $100$\,mK, but explains why the SQUID noise saturates at about $500$\,mK. 

The power dissipated in the resistors has to be transferred to the phonons in the SQUID chip at temperature $T_a$ and eventually to the experimental stage at temperature $T$ through a thermal resistance $R_a$, as shown in Fig.~\ref{thermalscheme}. Therefore, the SQUID chip will be also overheated, so that $T_a>T$. We expect the dominant thermal resistance from the chip to the refrigerator to be a Kapitza contact thermal resistance at the interface between the silicon crystal chip and the PC board on which it is glued by GE Varnish. Kapitza resistance can be usually written as $R_a=\frac{1}{c_K S} T^{-3}$, where $c_K$ is a constant, and $S$ is the contact area \cite{pobell}. Under this assumption, we expect Eq.~(4) to hold with $n=4$ and the crossover temperature given by \cite{hotelectron}:
\begin{equation}
  T_{co}=\left( \frac{4 W}{c_K A} \right)^\frac{1}{4}
\end{equation}
By assuming $T_{co} \approx 85$ mK as inferred from our data in Fig.~3, and taking into account the contact area $S\approx 10$ mm$^2$, we estimate $c_K \approx 4$\,W/m$^2$K$^4$, which is a realistic value \cite{pobell}. For comparison, in Ref.~\cite{wellstood} the value of $c_K=20$\,W/m$^2$K$^4$ is reported for the interface between silicon crystal and liquid helium.

While these considerations do not allow to identify the dissipation mechanism, they strongly support an interpretation of the observed saturation as a thermal effect, and provide a further justification for excluding the points below the crossover temperature for testing the CSL model.   

\end{document}